\documentclass[12pt]{article}%
\usepackage{amsmath,latexsym}
\usepackage{graphicx}
\usepackage{amsmath}
\usepackage{amsfonts}
\usepackage{amssymb}%
\begin{document}

\title{{Noncommutative-geometry inspired charged wormholes
   with low tidal forces}}
   \author{
Peter K.F. Kuhfittig* and Vance D. Gladney*\\
\footnote{kuhfitti@msoe.edu}
 \small Department of Mathematics, Milwaukee School of
Engineering,\\
\small Milwaukee, Wisconsin 53202-3109, USA}

\date{}
 \maketitle

\begin{abstract}\noindent
When Morris and Thorne first proposed that
wormholes might be actual physical structures
suitable for interstellar travel, they needed
to pay close attention to certain traversability
conditions such as low tidal forces, which
placed severe constraints on the wormhole
geometry.  Even more problematical was the
need for ``exotic matter" resulting from the
unavoidable violation of the null energy
condition required to hold a wormhole open.
The purpose of this paper is to overcome
these problems by starting with the charged
wormhole model of Kim and Lee and assuming a
noncommutative-geometry background: the
violation of the null energy condition can
be attributed to the latter, while the
electric charge allows the reduction of
the tidal forces to acceptable levels without
invoking the trivial zero-tidal-force
assumption. \\

\noindent
Keywords\\
Charged Wormholes, Noncommutative Geometry,
    Tidal Constraints\\

\end{abstract}

\section{Introduction}\label{S:Introduction}

Wormholes are handles or tunnels in
spacetime that link widely separated
regions of our Universe or different universes
altogether.  Morris and Thorne \cite{MT88}
proposed the following line element for the
wormhole spacetime:
\begin{equation}\label{E:line1}
ds^{2}=-e^{2\Phi(r)}dt^{2}+\frac{dr^2}{1-b(r)/r}
+r^{2}(d\theta^{2}+\text{sin}^{2}\theta\,
d\phi^{2}),
\end{equation}
using units in which $c=G=1$.  In this line
element, $b=b(r)$ is called the \emph{shape
function} and $\Phi=\Phi(r)$ is called the
\emph{redshift function}, which must be
everywhere finite to avoid an event horizon.
For the shape function we must have
$b(r_0)=r_0$, where $r=r_0$ is the radius of
the \emph{throat} of the wormhole.  An important
requirement is the \emph{flare-out condition}
at the throat: $b'(r_0)<1$, while $b(r)<r$
near the throat.  The flare-out condition can
only be met by violating the null energy
condition
\begin{equation}
   T_{\alpha\beta}k^{\alpha}k^{\beta}\ge 0
\end{equation}
for all null vectors $k^{\alpha}$, where
$T_{\alpha\beta}$ is the stress-energy tensor.
Matter that violates this condition is referred to
as ``exotic" in Ref. \cite{MT88}.  In particular,
for the radial outgoing null vector $(1,1,0,0)$,
the violation reads
$T_{\alpha\beta}k^{\alpha}k^{\beta}=\rho+
p_r<0$.  Here $T^t_{\phantom{tt}t}=-\rho(r)$,
the energy density, and $T^r_{\phantom{rr}r}=
p_r(r)$, the radial pressure.
($T^\theta_{\phantom{\theta\theta}\theta}=
T^\phi_{\phantom{\phi\phi}\phi}=p_t(r)$,
the lateral pressure.)

In Ref. \cite{MT88}, Morris and Thorne also
discussed the tidal forces that a traveler
would be subjected to and subsequently put
the following constraint on the redshift
function: $\Phi'(r)\le g_{\oplus}/
(c^2\sqrt{1-b(r)/r})$.  The feasibility of
this assumption will be discussed later.

The purpose of this paper is to revisit
these requirements by starting with the
modified charged wormhole model due to
Kim and Lee \cite{KL01} and assuming a
noncommutative-geometry background.

Similar issues involving traversable
wormholes in Lovelock gravity are discussed
in Refs. \cite{DD09, ZLP15, MZL15a}, with
emphasis on the energy conditions.  In
particular, Ref. \cite{MZL15a} considers
higher-dimensional thin-shell wormholes.
Einstein-Gauss-Bonnet wormholes satisfying
the weak energy condition are studied in
Ref. \cite{MZL15b}, while higher-dimensional
evolving wormholes satisfying the null energy
condition are discussed in Ref. \cite{ZLR14}.
Regarding the tidal constraints, it was noted
in Ref. \cite{pK15} that the zero-tidal-force
assumption is incompatible with quantum field
theory in classical general relativity.

\section{Noncommutative geometry}
An important outcome of string theory is
the realization that coordinates may become
noncommutative operators on a $D$-brane
\cite{eW96, SW99}.  The result is a
fundamental discretization of spacetime
due to the commutator
$[\textbf{x}^{\mu},\textbf{x}^{\nu}]
=i\theta^{\mu\nu}$, where $\theta^{\mu\nu}$ is
an antisymmetric matrix.  According to Refs.
\cite{SS03, NSS06, mR11}, noncommutativity
replaces point-like objects by smeared
objects, thereby  eliminating the divergences
that normally appear in general relativity.

Applications of noncommutative geometry are
numerous and varied.  One of the first was to
wormholes in semi-classical gravity by
Garattini and Lobo \cite{GL09}.  The same
authors analyzed the stability of gravastars
\cite{LG13}.  Gravitational lensing of
wormholes in noncommutative geometry was
studied in Ref. \cite{pK16a}, while the
search for higher-dimensional wormholes
was proposed in Ref. \cite{fR12}.

It seems natural to model the smearing by
means of a Gaussian distribution of minimal
length $\sqrt{\alpha}$ instead of the Dirac
delta function \cite{NSS06,mR11, fR12, pK13}.
A simpler but equally effective way is to
assume that the energy density of the static
and spherically symmetric and particle-like
gravitational source has the form
\begin{equation}\label{E:rho}
  \rho(r)=\frac{M\sqrt{\alpha}}
     {\pi^2(r^2+\alpha)^2};
\end{equation}
(see Refs. \cite{LL12} and \cite{NM08}.)
The point is that the mass $M$ is diffused
throughout the region of linear dimension
$\sqrt{\alpha}$ due to the uncertainty.
It should be noted that noncommutative
geometry is an intrinsic property of spacetime
and does not depend on particular features
such as curvature.

Next, we note that the Einstein field equations
$G_{\mu\nu}=
8\pi T_{\mu\nu}$ result in the
following forms:
\begin{equation}\label{E:Einstein1}
  \rho(r)=\frac{b'}{8\pi r^2},
\end{equation}
\begin{equation}\label{E:Einstein2}
   p_r(r)=\frac{1}{8\pi}\left[-\frac{b}{r^3}+
   2\left(1-\frac{b}{r}\right)\frac{\Phi'}{r}
   \right],
\end{equation}
and
\begin{equation}\label{E:Einstein3}
   p_t(r)=\frac{1}{8\pi}\left(1-\frac{b}{r}\right)
   \left[\Phi''-\frac{b'r-b}{2r(r-b)}\Phi'
   +(\Phi')^2+\frac{\Phi'}{r}-
   \frac{b'r-b}{2r^2(r-b)}\right].
\end{equation}
Here $p_r(r)$ is the radial pressure and $p_t(r)$
is the lateral pressure.
The conservation law
$T^{\mu\nu}_{\phantom{\mu\nu};\nu}=0$
implies that only two of Eqs.
(\ref{E:Einstein1})-(\ref{E:Einstein3}) are
independent.  So Eq. (\ref{E:Einstein3}) can be
obtained from the other two.  (See Ref. \cite{sS05}
for further details.)

As a final comment, to make use of Eq.
(\ref{E:rho}), we can keep the standard form
of the Einstein field equations in the sense
that the Einstein tensor retains its original
form, but the stress-energy tensor is modified
\cite{NSS06}.  It follows that the length
scale used need not be restricted to the
Planck scale.

\section{The modified Kim-Lee model}

Since black holes can carry an electric charge,
it is natural to assume that wormholes can do
likewise.  So it was proposed by Kim and Lee
\cite{KL01} that for a wormhole with constant
electric charge $Q$, the Einstein field
equations take on the form
\begin{equation}\label{E:EFE}
   G^{(0)}_{\mu\nu}+G^{(1)}_{\mu\nu}=8\pi
      [T^{(0)}_{\mu\nu}+T^{(1)}_{\mu\nu}].
\end{equation}
Given that the usual form is $G^{(0)}_{\mu\nu}=
8\pi T ^{(0)}_{\mu\nu}$, the modified form in
Eq. (\ref{E:EFE}) is obtained by adding the matter
term $T^{(1)}_{\mu\nu}$ to the right side  and
the corresponding back reaction term
$G^{(1)}_{\mu\nu}$ to the left side.  The proposed
metric is
\begin{equation}\label{E:line2}
  ds^2=-\left(1+\frac{Q^2}{r^2}\right)dt^2
   +\left(1-\frac{b(r)}{r}+\frac{Q^2}{r^2}\right)^{-1}dr^2\\
    +r^2(d\theta^2+\text{sin}^2\theta\,d\phi^2).
\end{equation}
Observe that whenever $b\equiv 0$, the wormhole
becomes a Reissner-Nordstr\"{o}m black hole,
and if $Q=0$, the spacetime reverts to a Morris-Thorne
wormhole, where the shape function $b=b(r)$  meets
the usual requirements.  Kim and Lee go on to
show that the metric, Eq. (\ref{E:line2}), is a
self-consistent solution of the Einstein field
equations.  The shape function $b=b(r)$ of the
Morris-Thorne wormhole is now replaced by the
effective shape function
\begin{equation}\label{E:eff}
    b_{\text{eff}}(r)=b(r)-\frac{Q^2}{r},
\end{equation}
to be discussed later.

To overcome certain difficulties in determining
the tidal constraints, we will use the following
modified metric already introduced in Refs.
\cite{pK11} and \cite{pK16}:
\begin{equation}\label{E:line3}
  ds^2=-\left(1+R(r)+\frac{Q^2}{r^2}\right)dt^2\\
   +\left(1-\frac{b(r)}{r}
   +\frac{Q^2}{r^2}\right)^{-1}dr^2
    +r^2(d\theta^2+\text{sin}^2\theta\,d\phi^2).
\end{equation}
here $S(r)$ is a differentiable function such
that $S(r)>0$ and $S'(r)>0$.  As noted in Ref.
\cite{pK11}, line element (\ref{E:line3})
remains a valid solution of the Einstein field
equations.

\section{The modified Kim-Lee model in noncommutative
     geometry}
In the Kim-Lee model, Eq. (\ref{E:line2}),
and in the modified model, Eq. (\ref{E:line3}),
\begin{equation}
   \text{total charge}=\int\int\int_V\rho_q(r)dV,
\end{equation}
where $\rho_q(r)$ is the charge density.
Adapting this to the noncommutative-geometry
background, we assume that the charge
density has the form
\begin{equation}\label{E:charged}
  \rho_q(r)=\frac{Q^2\sqrt{\alpha}}
     {\pi^2(r^2+\alpha)^2},
\end{equation}
where $Q^2$ refers to the Kim-Lee model.  To
obtain the smeared charge $Q^2_{\alpha}(r)$,
we evaluate
\begin{equation}\label{E:Q}
   Q^2_{\alpha}(r)=\int^r_{r_0}4\pi (r')^2
   \frac{Q^2\sqrt{\alpha}}
     {\pi^2[(r')^2+\alpha]^2}dr'
   =\frac{2Q^2\sqrt{\alpha}}{\pi}
   \left(\frac{1}{\sqrt{\alpha}}\text{tan}^{-1}
   \frac{r}{\sqrt{\alpha}}-
   \frac{r}{r^2+\alpha}\right).
\end{equation}
Observe that $Q^2_{\alpha}(r_0)=0$; also,
seen from a large distance, $Q^2_{\alpha}(r)$
becomes $Q^2$.

Similarly, from Eqs. (\ref{E:Einstein1}) and
(\ref{E:eff}), we get for the effective shape
function,
\begin{multline}\label{E:shape1}
  b_{\text{eff}}(r)=\frac{4M\sqrt{\alpha}}{\pi}
  \left(\frac{1}{\sqrt{\alpha}}\text{tan}^{-1}
  \frac{r}{\sqrt{\alpha}}-\frac{r}{r^2+\alpha}
  \right)\\
  -\frac{1}{r}\frac{2Q^2\sqrt{\alpha}}
  {\pi}\left(\frac{1}{\sqrt{\alpha}}\text{tan}^{-1}
  \frac{r}{\sqrt{\alpha}}-\frac{r}{r^2+\alpha}
  \right)+C.
\end{multline}
Since $Q^2_{\alpha}(r_0)=0$, we obtain from
$b(r_0)=r_0$,
\begin{equation}\label{E:shape2}
  b_{\text{eff}}(r_0)=\frac{4M\sqrt{\alpha}}{\pi}
  \left(\frac{1}{\sqrt{\alpha}}\text{tan}^{-1}
  \frac{r_0}{\sqrt{\alpha}}-\frac{r_0}{r_0^2+\alpha}
  \right)-0+C=r_0
\end{equation}
and hence the constant $C$.  Thus
\begin{multline}\label{E:shape3}
  b_{\text{eff}}(r)=\frac{4M\sqrt{\alpha}}{\pi}
  \left(\frac{1}{\sqrt{\alpha}}\text{tan}^{-1}
  \frac{r}{\sqrt{\alpha}}-\frac{r}{r^2+\alpha}
  \right)\\
  -\frac{1}{r}\frac{2Q^2\sqrt{\alpha}}
  {\pi}\left(\frac{1}{\sqrt{\alpha}}\text{tan}^{-1}
  \frac{r}{\sqrt{\alpha}}-\frac{r}{r^2+\alpha}
  \right)
  -\frac{4M\sqrt{\alpha}}{\pi}
  \left(\frac{1}{\sqrt{\alpha}}\text{tan}^{-1}
  \frac{r_0}{\sqrt{\alpha}}-\frac{r_0}{r_0^2+\alpha}
  \right)+r_0.
\end{multline}
For a wormhole solution, we require that
$b'_{\text{eff}}(r_0)>0$ and
$b'_{\text{eff}}(r_0)<1$, the flare-out
condition mentioned earlier.  To that end,
we determine
\begin{equation}\label{E:derivative}
  b'_{\text{eff}}(r_0)=\frac{4M\sqrt{\alpha}}
  {\pi}\frac{2r_0^2}{(r_0^2+\alpha)^2}-
  \frac{1}{r_0}\frac{2Q^2\sqrt{\alpha}}{\pi}
  \frac{2r_0^2}{(r_0^2+\alpha)^2}=
  \frac{4r_0^2\sqrt{\alpha}}
  {\pi (r_0^2+\alpha)^2}
     \left(2M-\frac{1}{r_0}Q^2\right)
\end{equation}
since $Q^2_{\alpha}(r_0)=0$.  Looking
ahead to the traversability conditions
which involve measurements in meters, we will
assume that $r_0>1$ m.  So to keep
$b'_{\text{eff}}(r_0)$ positive, we must have
\begin{equation}
    Q^2\le 2M.
\end{equation}
Since $\sqrt{\alpha}\ll M$, it also follows
that
\begin{equation}\label{E:flare}
   b'_{\text{eff}}(r_0)<1,
\end{equation}
as required.  Moreover, $b_{\text{eff}}(r)
<r$ near the throat and
\[
   \text{lim}_{r\rightarrow\infty}
  \frac{b_{\text{eff}}(r)}{r}=0,
\]
so that the wormhole spacetime is
asymptotically flat.

From the exoticity condition in Ref.
\cite{MT88}, $b'_{\text{eff}}(r_0)<1$
is equivalent to the violation of the null
energy condition, requiring the need for
``exotic matter."  According to Eqs.
(\ref{E:derivative}) and (\ref{E:flare}),
however, this violation is simply a consequence
of the noncommutative-geometry background.

To study the effect that the tidal forces
may have on the traveler, we return to the
criterion
\begin{equation}\label{E:Phiprime1}
    \Phi'(r)\le\frac{g_{\oplus}}
    {c^2\sqrt{1-b(r)/r}}
\end{equation}
mentioned in Sec. \ref{S:Introduction}.  At
the throat of the wormhole, this condition
is trivially satisfied.  It is subsequently
proposed in Ref. \cite{MT88} that the space
stations be far enough away from the throat
so that $b(r)/r$ becomes negligible.  The
result is the requirement
\begin{equation}\label{E:Phiprime2}
   \Phi'\lesssim (9.2\times 10^{15}\,\text{m})
   ^{-1}\approx (10^{16}\,\text{m})^{-1}.
\end{equation}
In the modified Kim-Lee model, Eq.
(\ref{E:line3}),
\[
    e^{2\Phi}=1+R(r)+\frac{Q^2}{r^2}
\]
yields
\begin{equation}
   \Phi(r)=\frac{1}{2}\text{ln}
   \left[1+R(r)+\frac{1}{r^2}
   \frac{2Q^2\sqrt{\alpha}}{\pi}
   \left(\frac{1}{\sqrt{\alpha}}
   \text{tan}^{-1}\frac{r}{\sqrt{\alpha}}
   -\frac{r}{r^2+\alpha}
   \right)\right]
\end{equation}
and
\begin{equation}\label{E:Phiprime3}
   \Phi'(r)=\frac{\frac{1}{2}R'(r)
   -\frac{1}{r^3}\frac{2Q^2\sqrt{\alpha}}
   {\pi}\left(\frac{1}{\sqrt{\alpha}}
   \text{tan}^{-1}\frac{r}{\sqrt{\alpha}}
   -\frac{r}{r^2+\alpha}\right)
   +\frac{2Q^2\sqrt{\alpha}}
   {\pi (r^2+\alpha)^2}}
   {1+R(r)+\frac{1}{r^2}
   \frac{2Q^2\sqrt{\alpha}}{\pi}
   \left(\frac{1}{\sqrt{\alpha}}
   \text{tan}^{-1}
   \frac{r}{\sqrt{\alpha}}
   -\frac{r}{r^2+\alpha}
   \right)}.
\end{equation}
As noted earlier, $R(r)>0$ and $R'(r)>0$,
but it quickly becomes apparent that in view
of Inequality (\ref{E:Phiprime2}), $R(r)$
has to be fine-tuned so that
\begin{equation}\label{E:finetune}
    \frac{1}{2}R'(r)-\frac{1}{r^3}
    \frac{2Q^2\sqrt{\alpha}}{\pi}
    \left(\frac{1}{\sqrt{\alpha}}\text{tan}^{-1}
    \frac{r}{\sqrt{\alpha}}
    -\frac{r}{r^2+\alpha}\right)
\end{equation}
becomes negligibly small near the proposed
location of the station.

\emph{Remark:} While the need for fine-tuning
may not be desirable, it does not present a
serious conceptual problem here: for large $r$,
expression (\ref{E:finetune}) approaches
\[
     \frac{1}{2}R'(r)-\frac{Q^2}{r^3}+
     \frac{2Q^2\sqrt{\alpha}}
        {\pi r^2(r^2+\alpha)}.
\]
So $\frac{1}{2}R'\approx Q^2/r^3$ near the station
where $R(r)$ has the approximate form $-Q^2/r^2
+C$, $C>0$; $C$ can be chosen so that the
denominator in Eq. (\ref{E:Phiprime3})
exceeds unity.  As a result
\[
   \Phi'(r)\lesssim \frac{2Q^2\sqrt{\alpha}}
       {\pi (r^2+\alpha)^2}.
\]
Eqs. (\ref{E:Phiprime2}) and (\ref{E:Phiprime3})
now yield the condition
\begin{equation}\label{E:tidal}
 \ \Phi'(r)\lesssim\frac{2Q^2\sqrt{\alpha}}
  {\pi (r^2+\alpha)^2}<(10^{16}\,\text{m})^{-1}.
\end{equation}

This requirement places the following constraint
on the parameter $\alpha$:
\begin{equation}\label{E:constraint}
 \sqrt{\alpha}<\frac{(r^2+\alpha)^2\times 10^{-16}}
 {(2Q^2)/\pi}\,\text{m}.
\end{equation}
According to Ref. \cite{NSS06}, noncommutativity
is not visible at presently accessible energies
if $\sqrt{\alpha}<10^{-18}$ m.  So according to
Inequality (\ref{E:constraint}), the constraint
on $\alpha$ is much less severe and so would in
principle be observable.

\section{Conclusion}
While Morris and Thorne \cite{MT88} had demonstrated
that wormholes could be macroscopic structures
suitable for interstellar travel, they could not
avoid certain problems with traversability.  In
particular, the structures would have to be
designed to produce low tidal forces, while
the unavoidable violation of the null energy
condition called for the use of ``exotic matter."
The latter requirement is not only difficult to
meet classically, but may even be unphysical.
The former was met by assuming a constant
redshift function, so that $\Phi'\equiv 0$,
called the ``zero-tidal-force solution."

To overcome these problems, this paper
discusses the modified wormhole model due
to Kim and Lee in conjunction with a
noncommutative-geometry background.  The
violation of the null energy condition can
be attributed to the latter, thereby avoiding
the need for exotic matter.  The presence of
$Q^2$ in the modified Kim-Lee model allows
the reduction of the tidal forces to
acceptable levels without invoking the
trivial zero-tidal-force assumption, again
made possible by the noncommutative-geometry
background.  Finally, the condition on
$\Phi'$ places a constraint on the value
of $\sqrt{\alpha}$, but one that is well
above the limit of observability.


\begin{thebibliography}{20}

\bibitem{MT88}Morris, M.S. and Thorne, K.S. (1988)
   Wormholes in spacetime and their use for
   interstellar travel: A tool for teaching
   general relativity. \emph{American Journal of
   Physics}, \textbf{56}, 395-412.
\bibitem{KL01}Kim, S.-W. and Lee, H. (2001) Exact
   solutions of charged wormhole.
   \emph{ Physical Review D}, \textbf{63},
   Article ID: 064014.
\bibitem{DD09}Dehghani, M.H. and Dayyani, Z. (2009)
   Lorentzian wormholes in Lovelock gravity.
   \emph{Physical Revue D}, \textbf{79},
   Article ID: 064010.
\bibitem{ZLP15}Zangeneh, M.K., Lobo, F.S.N. and
   Dehghani, M.H. (2015) Traversable wormholes
   satisfying the weak energy condition in third-order
   Lovelock gravity. \emph{Physical Review D},
   \textbf{92}, Article ID: 124049.
\bibitem{MZL15a}Mehdizadeh, M.R., Zangeneh, M.K. and
   Lobo, F.S.N. (2015) Higher-dimensional thin-shell
   wormholes in third-order Lovelock gravity.
   \emph{Physical Review D}, \textbf{92},
   Article ID: 044022.
\bibitem{MZL15b}Mehdizadeh, M.R., Zangeneh, M.K. and
   Lobo, F.S.N. (2015) Einstein-Gauss-Bonnet traversable
   wormholes satisfying the weak energy condition.
   \emph{Physical Review D}. \textbf{91},
   Article ID: 084004.
\bibitem{ZLR14}Zangeneh, M.K., Lobo, F.S.N. and Riazi,
   N. (2014) Higher-dimensional evolving wormholes
   satisfying the null energy condition.
   \emph{Physical Review D}, \textbf{90},
   Article ID: 024072.
\bibitem{pK15}Kuhfittig, P.K.F. (2015) Macroscopic
   traversable wormholes with zero tidal forces inspired
   by noncommutative geometry. \emph{International
   Journal of Modern Physics D}, \textbf{24},
   Article ID: 1550023.
\bibitem{eW96}Witten, E. (1996) Bound states of
   strings and $p$-branes. \emph{Nuclear Physics B},
    \textbf{460}, 335-350.
\bibitem{SW99}Seiberg, N. and Witten, E. (1999)
   String theory and noncommutative geometry.
   \emph{Journal of High Energy Physics},
    \textbf{9909}, Article ID: 032.
\bibitem{SS03}Smailagic, A. and Spallucci, E. (2003)
   Feynman path integral on the non-commutative plane.
   \emph{Journal of Physics A}, \textbf{36}, L-467-L-471.
\bibitem{NSS06}Nicolini, P., Smailagic, A. and Spallucci,
    E. (2006) Noncommutative geometry inspired
    Schwarzschild black hole. \emph{Physics Letters B},
    \textbf{632}, 547-551.
\bibitem{mR11}Rinaldi, M. (2011) A new approach to
   non-commutative inflation. \emph{Classical and
   Quantum Gravity}, \textbf{28}, Article ID: 105022.
\bibitem{GL09}Garattini, R. and Lobo, F.S.N. (2009)
   Self-sustained traversable wormholes in
   noncommutative geometry. \emph{Physics Letters B},
   \textbf{71}, 146-152.
\bibitem{LG13}Lobo, F.S.N. and Garattini, R. (2013)
   Linearized stability analysis of gravastars in
   noncommutative geometry. \emph{Journal of High Energy
   Physics}, \textbf{1312}, Article ID: 065.
\bibitem{pK16a}Kuhfittig, P.K.F. (2016) Gravitational
   lensing of wormholes in noncommutative geometry.
   \emph{Scientific Voyage}, \textbf{2}, 1-8
\bibitem{fR12}Rahaman, F., Kuhfittig, P.K.F., Ray, S.
   and Islam, S. (2012) Searching for higher dimensional
   wormholes with noncommutative geometry. \emph{Physical
   Review D}, \textbf{86}, Article ID: 106101.
\bibitem{pK13}Kuhfittig, P.K.F. (2013) Macroscopic
   wormholes in noncommutative geometry. \emph{
   International Journal of Pure and Applied
   Mathematics}, \textbf{89}, 401-408.
\bibitem{LL12}Liang, J. and Liu, B. (2012)
   Thermodynamics of noncommutative geometry inspired
   BTZ black holes based on Lorentzian smeared mass
   distribution. \emph{Europhysics Letters}, \textbf{100},
   Article ID: 30001.
\bibitem{NM08}Nozari, K. and Mehdipour, S.H. (2008)
   Hawking radiation as quantum tunneling for a
   noncommutative Schwarzschild black hole.
   \emph{Classical and Quantum Gravity}, \textbf{25},
   Article ID: 175015.
\bibitem{sS05}Sushkov, S.V. (2005) Wormholes supported
   by a phantom energy. \emph{Physical Review D},
   \textbf{71}, Article ID: 043520.
\bibitem{pK11}Kuhfittig, P.K.F. (2011) On the
   feasibility of charged wormholes. \emph{Central
   European Journal of Physics}, \textbf{9}, 1144-1150.
\bibitem{pK16}Kuhfittig, P.K.F. (2016) The effect of
   conformal symmetry on charged wormholes. \emph{
   Journal of Applied Mathematics and Physics},
    \textbf{4}, 2117-2125.
 \end{thebibliography}
\end{document}